\documentclass[aps,prl,
              reprint,
              amsmath,amssymb,amsfonts,groupedaddress,
              longbibliography,
              superscriptaddress,floatfix]{revtex4-1}

\usepackage{pdfpages}

\makeatletter
\AtBeginDocument{\let\LS@rot\@undefined}
\makeatother
              
\usepackage{graphicx}
\usepackage{dcolumn}
\usepackage{bm}
\usepackage{amsbsy}
\usepackage{units}

\usepackage{color}

\usepackage[breaklinks=true]{hyperref}
\usepackage{hyperref}
  \hypersetup{
  colorlinks=true,
  linkcolor=red,
  citecolor=green,
  filecolor=black,
  menucolor=black,
  urlcolor=blue
  }
 
\usepackage{breakurl}

\newcommand{\sdist}{\kern 0.20em}
\renewcommand{\eqref}[1]{Eq.\sdist(\ref{#1})}
\newcommand{\figref}[1]{Fig.\sdist\ref{#1}}

\allowdisplaybreaks

\usepackage{upgreek}

\begin{document}

\title{Extremely Dense Gamma-Ray Pulses in Electron Beam-Multifoil Collisions}

\author{Archana Sampath}
\affiliation{Max-Planck-Institut f\"ur Kernphysik, Saupfercheckweg 1, D-69117 Heidelberg, Germany}
\author{Xavier Davoine}
\affiliation{CEA, DAM, DIF, 91297 Arpajon, France}
\affiliation{Universit\'{e} Paris-Saclay, CEA, LMCE, 91680 Bruy\`{e}res-le-Ch\^{a}tel, France}
\author{S\'{e}bastien Corde}
\affiliation{LOA, ENSTA Paris, CNRS, Ecole Polytechnique, Institut Polytechnique de Paris, 91762 Palaiseau, France}
\author{Laurent Gremillet}
\affiliation{CEA, DAM, DIF, 91297 Arpajon, France}
\affiliation{Universit\'{e} Paris-Saclay, CEA, LMCE, 91680 Bruy\`{e}res-le-Ch\^{a}tel, France}
\author{Max Gilljohann}
\affiliation{LOA, ENSTA Paris, CNRS, Ecole Polytechnique, Institut Polytechnique de Paris, 91762 Palaiseau, France}
\author{Maitreyi Sangal}
\affiliation{Max-Planck-Institut f\"ur Kernphysik, Saupfercheckweg 1, D-69117 Heidelberg, Germany}
\author{Christoph H. Keitel}
\affiliation{Max-Planck-Institut f\"ur Kernphysik, Saupfercheckweg 1, D-69117 Heidelberg, Germany}
\author{Robert Ariniello}
\affiliation{University of Colorado Boulder, Department of Physics, Center for Integrated Plasma Studies, Boulder, Colorado 80309, USA}
\author{John Cary}
\affiliation{University of Colorado Boulder, Department of Physics, Center for Integrated Plasma Studies, Boulder, Colorado 80309, USA}
\author{Henrik Ekerfelt}
\affiliation{SLAC National Accelerator Laboratory, Menlo Park, CA 94025, USA}
\author{Claudio Emma}
\affiliation{SLAC National Accelerator Laboratory, Menlo Park, CA 94025, USA}
\author{Frederico Fiuza}
\affiliation{SLAC National Accelerator Laboratory, Menlo Park, CA 94025, USA}
\author{Hiroki Fujii}
\affiliation{University of California Los Angeles, Los Angeles, CA 90095, USA}
\author{Mark Hogan}
\affiliation{SLAC National Accelerator Laboratory, Menlo Park, CA 94025, USA}
\author{Chan Joshi}
\affiliation{University of California Los Angeles, Los Angeles, CA 90095, USA}
\author{Alexander Knetsch}
\affiliation{LOA, ENSTA Paris, CNRS, Ecole Polytechnique, Institut Polytechnique de Paris, 91762 Palaiseau, France}
\author{Olena Kononenko}
\affiliation{LOA, ENSTA Paris, CNRS, Ecole Polytechnique, Institut Polytechnique de Paris, 91762 Palaiseau, France}
\author{Valentina Lee}
\affiliation{University of Colorado Boulder, Department of Physics, Center for Integrated Plasma Studies, Boulder, Colorado 80309, USA}
\author{Mike Litos}
\affiliation{University of Colorado Boulder, Department of Physics, Center for Integrated Plasma Studies, Boulder, Colorado 80309, USA}
\author{Kenneth Marsh}
\affiliation{University of California Los Angeles, Los Angeles, CA 90095, USA}
\author{Zan Nie}
\affiliation{University of California Los Angeles, Los Angeles, CA 90095, USA}
\author{Brendan O'Shea}
\affiliation{SLAC National Accelerator Laboratory, Menlo Park, CA 94025, USA}
\author{J. Ryan Peterson}
\affiliation{SLAC National Accelerator Laboratory, Menlo Park, CA 94025, USA}
\affiliation{Stanford University, Physics Department, Stanford, CA 94305, USA}
\author{Pablo San Miguel Claveria}
\affiliation{LOA, ENSTA Paris, CNRS, Ecole Polytechnique, Institut Polytechnique de Paris, 91762 Palaiseau, France}
\author{Doug Storey}
\affiliation{SLAC National Accelerator Laboratory, Menlo Park, CA 94025, USA}
\author{Yipeng Wu}
\affiliation{University of California Los Angeles, Los Angeles, CA 90095, USA}
\author{Xinlu Xu}
\affiliation{SLAC National Accelerator Laboratory, Menlo Park, CA 94025, USA}
\author{Chaojie Zhang}
\affiliation{University of California Los Angeles, Los Angeles, CA 90095, USA}
\author{Matteo Tamburini}\email{matteo.tamburini@mpi-hd.mpg.de}
\affiliation{Max-Planck-Institut f\"ur Kernphysik, Saupfercheckweg 1, D-69117 Heidelberg, Germany}

\date{\today}

\begin{abstract}
Sources of high-energy photons have important applications in almost all areas of research. However, the photon flux and intensity of existing sources is strongly limited for photon energies above a few hundred keV. Here we show that a high-current ultrarelativistic electron beam interacting with multiple submicrometer-thick conducting foils can undergo strong self-focusing accompanied by efficient emission of gamma-ray synchrotron photons. Physically, self-focusing and high-energy photon emission originate from the beam interaction with the near-field transition radiation accompanying the beam-foil collision. This near field radiation is of amplitude comparable with the beam self-field, and can be strong enough that a single emitted photon can carry away a significant fraction of the emitting electron energy. After beam collision with multiple foils, femtosecond collimated electron and photon beams with number density exceeding that of a solid are obtained. The relative simplicity, unique properties, and high efficiency of this gamma-ray source open up new opportunities for both applied and fundamental research including laserless investigations of strong-field QED processes with a single electron beam.
\end{abstract}

\maketitle

The generation of high-energy, dense and collimated photon beams is of great interest both to fundamental and applied research. Indeed, such beams enable new avenues for research in strong-field QED, relativistic plasma astrophysics, and high-energy physics~\cite{mourouRMP06, marklundRMP06, ruffiniPR10, dipiazzaRMP12}. In particular, solid-density photon beams allow matterless photon-photon physics studies, where traditional schemes are limited in luminosity due to the low density of high-energy photons~\cite{telnovNIMA90}. A source of high-energy, solid-density photon beams also enables the generation of neutral collimated ultradense electron-positron jets, opening a unique portal to novel relativistic laboratory astrophysics studies~\cite{sarriNC15, chenPRL15, lobetPRL15, arrowsmithXXX20}. Moreover, energetic solid-density electron and photon beams make it possible to access important unexplored regimes in high-density beam physics~\cite{delgaudioPRL20, xuXXX20}. Furthermore, intense sources of high-energy photons have broad applications in industry, medicine, and materials science~\cite{bilderbackJPB05, ullrichARPC12, cordeRMP13, lightsources}.

The growing interest in intense high-energy photon sources has recently stimulated several proposals to further increase the attainable photon energy and flux. These proposals include high-power laser-plasma interactions~\cite{ridgersPRL12, nakamuraPRL12, jiPoP14, liPRL15, zhuNJP15, starkPRL16, changSR17, wangPNAS18, huangPPCF18, huangNJP19, vranicPOP19, jirkaSR20, ferriPRL18, zhuSA20}, plasma instabilities~\cite{benedettiNP18}, QED cascades~\cite{jirkaPRE16, tamburiniSR17}, multiple colliding laser pulses~\cite{gonoskovPRX17, magnussonPRL19} and beamstrahlung~\cite{yakimenkoPRL19, delgaudioPRAB19, tamburiniXXX19}. A number of experiments, where the generated photon beam properties could be accurately measured and tuned, were also successfully performed~\cite{schlenvoigtNP08, kneipNP10, cipicciaNP11, phuocNP12, sarriPRL14, yanNP17, colePRX18, poderPRX18}. In those schemes, however, the achievable density remains less than $\sim10^{24}\text{ m$^{-3}$}$.

\begin{figure}[tb]
\centering
\includegraphics[width=\linewidth]{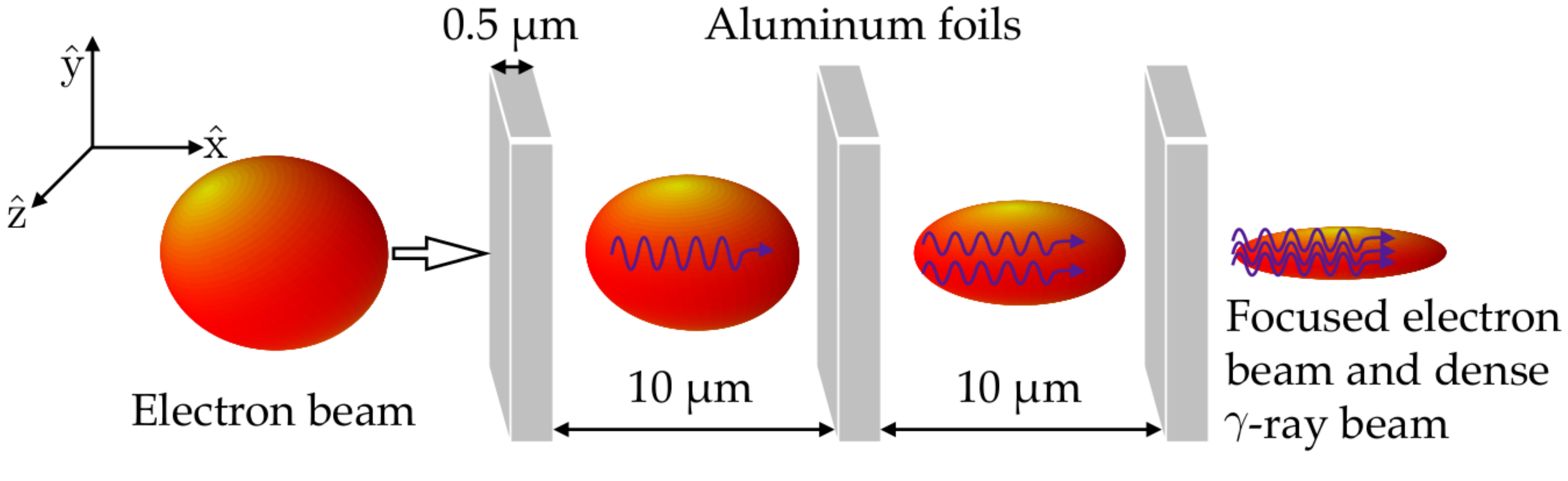}
\caption{Schematic setup. An ultrarelativistic electron beam sequentially collides with aluminum foils. At each beam-foil collision, a strong transverse force which focuses the electron beam and leads to copious gamma-ray emission is induced.}
\label{fig:1}
\end{figure}
\begin{figure*}[tb]
\centering
\includegraphics[width=\linewidth]{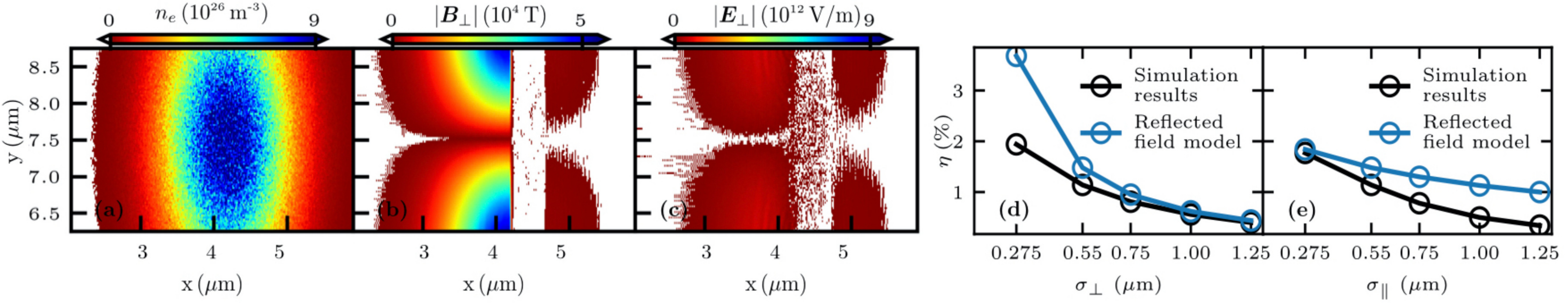} 
\caption{(a) Electron beam density, (b) transverse magnetic field, and (c) transverse electric field in the collision with a 0.5$\mu$m-thick aluminum foil. For comparison, the peak magnetic and electric beam self-fields are $3.1\times 10^4$~T, and $9.4\times10^{12}$~V/m, respectively. (d)~Electron beam to radiation energy conversion efficiency $\eta$ as a function of $\sigma_\perp$ in the collision with one foil. The electron beam has 2~nC charge, 10~GeV energy, and $\sigma_\parallel=0.55\text{ $\mu$m}$. Black circles: 3D PIC simulations results; blue circles: reflected-field model predictions. (e) Same as in panel~(d) but for $\sigma_\perp=0.55\text{ $\mu$m}$ and as a function of $\sigma_\parallel$.}
\label{fig:2}
\end{figure*}

Here we introduce a novel concept for an ultraintense gamma-ray source based on the interaction of a single high-current ultrarelativistic electron beam with multiple submicrometer-thick conducting foils (see \figref{fig:1}). By using fully 3D particle-in-cell (PIC) simulations, we show that: (i) An ultrarelativistic (10~GeV), dense ($4.7 \times 10^{27}\text{ m$^{-3}$}$) electron beam can be radially focused up to $5.5 \times 10^{29}\text{ m$^{-3}$}$, i.e., beyond the $1.8\times 10^{29}\text{ m$^{-3}$}$ electron density of solid aluminum; (ii) Electron beam focusing is accompanied by intense synchrotron photon emission with more than 30\% of the electron beam energy eventually converted into a $2.8 \times 10^{29}\text{ m$^{-3}$}$ peak density collimated gamma-ray beam (with a maximum density of $8.5 \times 10^{29}\text{ m$^{-3}$}$ achieved in the course of the interaction); (iii) When the electron beam density exceeds the foil electron density, the beam self-fields expel the target electrons and create an electron-depleted channel through the foil. The fields experienced by the beam electrons inside the plasma channel are so high that the quantum parameter $\chi \approx \gamma |\bm{f}_\perp|/e F_{\text{cr}}$ exceeds unity~\cite{ritusJSLR85, Baier-book}. Here $\bm{f}_\perp = q (\bm{E}_\perp + \bm{\beta} \times \bm{B}) $ is the Lorentz force transverse to the beam velocity, $\gamma$ the beam relativistic factor, $q=-e$ the electron charge, $F_{\text{cr}}=m_e^2 c^3/e \hbar \approx 1.3\times 10^{18}\,\text{V/m}$ the QED critical field~\cite{ritusJSLR85, Baier-book, dipiazzaRMP12}. This opens up the possibility of laserless strong-field QED investigations with only one ultrarelativistic electron beam~\cite{sampathThes}. 

We start by considering the free propagation of an electron beam in vacuum. The electric $\bm{E}$ and magnetic $\bm{B}$ self-fields of a cold electron beam in vacuum are related by $\bm{B} = \bm{\beta} \times \bm{E}$~\cite{jackson-book}, where $\bm{\beta}=\bm{v}/c$ is the normalized beam velocity (Gaussian units are employed for equations). Thus, $\bm{f}_\perp = q \bm{E}_\perp/\gamma^2$ is strongly suppressed for large $\gamma$, and the beam propagates almost ballistically over relatively long distances in vacuum.

When a beam collides with a conductor, it can be subject to strong near-field coherent transition radiation (CTR), which alters the nearly perfect cancellation of the electric and magnetic terms in the Lorentz force. Electromagnetic boundary conditions require that the electric field component tangential to the surface of a perfect conductor must be continuous and zero at the conductor surface, whereas the tangential magnetic field can be discontinuous and remains large~\cite{jackson-book}. Thus, when an electron beam encounters a conductor, the magnetic term of the Lorentz force, which drives beam focusing, can overcome the electric term, which drives beam expansion. Effectively, when the beam length is smaller than its transverse size, this process can be visualized as a beam colliding with its image charge (see below and Supplemental Material~\footnote{See Supplemental Material for (i) modeling of the near-field CTR; (ii) PIC simulations at presently achievable beam densities; (iii) a movie of the electron and gamma beam evolution in the interaction with 20 consecutive foils.} for details on the near-field CTR fields, which include Refs.\cite{Ginzburg_1990, Gradshteyn_2007, Verzilov_2000, Harvey_1979, Castellano_1998, Lifschitz2009}). Notice that a large $\bm{f}_\perp$ naturally results in intense emission of radiation. For instance, in the classical regime the radiated power (mean photon energy) is proportional to $\gamma^2 \bm{f}_\perp^2$ ($\gamma^2 \bm{f}_\perp$)~\cite{jackson-book, Baier-book}.

\begin{figure*}[tb]
\centering
\includegraphics[width=\linewidth]{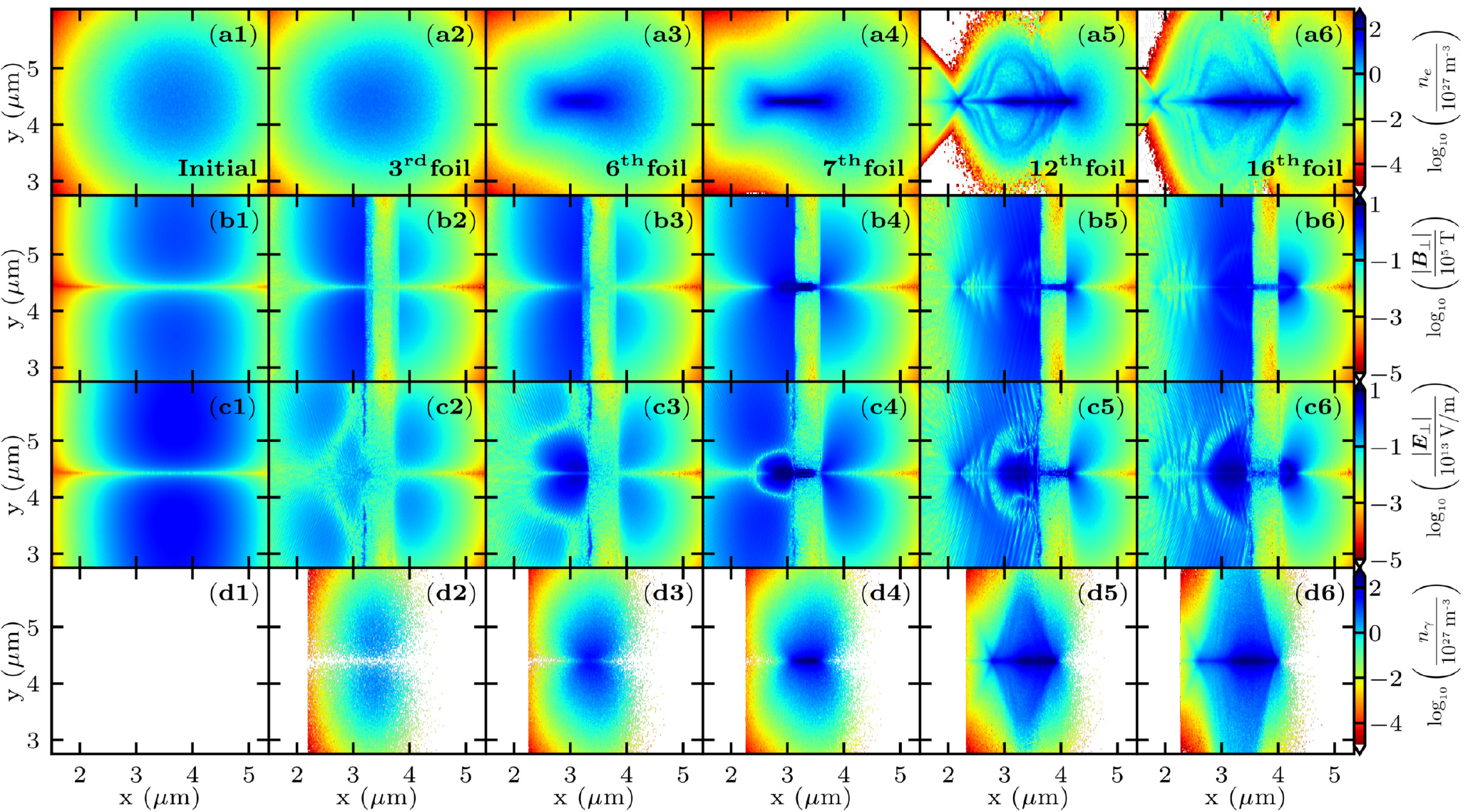} 
\caption{Beam evolution. First column, initial electron beam density~(a1), its magnetic~(b1) and electric~(c1) fields, and the initial photon density~(d1). Second to sixth column, same quantities as in the first column but at the 3rd (a2)-(d2), the 6th (a3)-(d3), the 7th (a4)-(d4), the 12th (a5)-(d5), and the 16th (a6)-(d6) beam-foil interaction, respectively.}
\label{fig:3}
\end{figure*}

For modeling, we consider an ultrarelativistic cold electron beam with cylindrical symmetry around its propagation axis $x$. The description is simplified by employing cylindrical coordinates with $r=\sqrt{y^2+z^2}$, $\theta=\arctan(z/y)$, and $x$ being the radial, azimuthal and vertical components, respectively. We assume that cylindrical symmetry is preserved throughout the interaction. Hence, fields are independent of $\theta$, the azimuthal electric field $E_\theta$ and the radial $B_r$ and vertical $B_x$ components of the magnetic field are zero. Here beam and conductor fields are denoted by the superscript $b$ and $c$, respectively. For an ultrarelativistic charge distribution $\rho(x,r,t) = \rho_0 e^{-r^2 / 2 \sigma_\perp^2} e^{-(x - x_0 - v t)^2 / 2 \sigma_\parallel^2}$ with $N_e$ electrons, initial position $x_0$, velocity $v$ along $x$, and peak charge density $\rho_0 = q N_e/(2\pi)^{3/2} \sigma_\perp^2 \sigma_\parallel$, $E_r^b \gg E_x^b \approx 0$,
\begin{equation} \label{eq:Er}
E_r^b (x,r,t) =  \frac{2 q N_e}{\sqrt{2 \pi} \sigma_\parallel r} \left(1-e^{-r^2/2 \sigma_\perp^2}\right) e^{-(x -x_0 - v t)^2/2\sigma_\parallel^2},
\end{equation}
and $B_\theta^b(x,r,t) = \beta E_r^b (x,r,t)$, provide an approximate solution to Maxwell equations up to terms of order $1/\gamma^2$ around the beam~\cite{sampathThes}. To evaluate $E_r^c (x,r,t)$ and $B_\theta^c(x,r,t)$, we consider a flat perfectly conducting foil with front surface at $x=0$. When the electron beam is outside the conductor, the method of images can be employed for determining $E_r^c (x,r,t)$ and $B_\theta^c(x,r,t)$ in $x<0$~\cite{hammondIEE60}. This method cannot be applied when the beam enters the foil, because the image would be located in $x<0$, where conductor fields must satisfy source-free Maxwell equations. However, when $\sigma_\perp \gg \sigma_\parallel$ one can approximate $E_r^c(x,r,t)$ and $B_\theta^c(x,r,t)$ with the image charge fields. This ``reflected-field'' approximation holds because CTR, which is emitted with transverse size $\sigma_\perp$ and typical wavelength $\sigma_\parallel$, undergoes weak diffraction over a Rayleigh length of approximately $\sigma_\perp^2/\sigma_\parallel \gg \sigma_\parallel$ from the boundary. The opposite limit $\sigma_\perp \ll \sigma_\parallel$, corresponds to the magnetostatic approximation, yielding a vanishing $B_\theta^c$ and a surface-localized $E_r^c$ (see Supplemental Material). Note that beam focusing in the $\sigma_\perp \ll \sigma_\parallel$ limit has been demonstrated in accelerators~\cite{adlerPA82, humphriesPA83, humphriesJAP88, humphriesAPL89, fernslerJAP90}.

The electron beam to radiated energy conversion efficiency $\eta$ can be calculated from $E_r=E_r^b+E_r^c$ and $B_\theta=B_\theta^b+B_\theta^c$, where \eqref{eq:Er} is employed for the beam and image charge fields. The average energy radiated per particle per unit time is conveniently approximated as~\cite{Baier-book} $\dot{\varepsilon}_\gamma = 2 \alpha m_e c^2 \chi^2 / 3 \tau_c [1 + 4.8 (1+\chi) \ln (1+1.7\chi) + 2.44 \chi^2]^{2/3}$, where $\alpha = e^2/\hbar c$ is the fine-structure constant, $\tau_c = \hbar / m_e c^2$ is the Compton time, and $\chi \approx \gamma |E_r - B_\theta | / F_{\text{cr}}$. Thus, 
\begin{equation} \label{eq:eta}
\eta = \frac{2 \pi \int_{-\infty}^{+\infty}{dt \int_{-\infty}^{0}{dx \int_{0}^{+\infty}{dr\, r \rho(x,r,t) \dot{\varepsilon}_\gamma[\chi(x,r,t)]}}}}{\gamma m_e c^2 q N_e}.
\end{equation}
In \eqref{eq:eta} we have assumed that all electrons have the same initial momentum and energy $\gamma m_e c^2$. Furthermore, we have neglected the change in $\gamma$ during the beam-foil interaction. The triple integral in \eqref{eq:eta} can be carried out numerically. 

Figure~\ref{fig:2} shows the results of 3D PIC simulations of a cold electron beam colliding with one 0.5$\mu$m-thick aluminum foil. The electron beam has 2~nC charge, 10~GeV energy, and Gaussian spatial distribution with $\sigma_\parallel=0.55\,\mu$m, $\sigma_\perp=1.25\,\mu$m, and $9.2 \times 10^{26}$~m$^{-3}$ density. Figure~\ref{fig:2}(a) displays a snapshot of the electron beam density when the beam center has reached the front surface of the foil. Figures~\ref{fig:2}(b) and \ref{fig:2}(c) show the transverse magnetic $\bm{B}_\perp$ and electric field $\bm{E}_\perp$, respectively. Whilst $\bm{B}_\perp$ is amplified and its peak value nearly doubles with respect to the beam self-field ($3.1\times 10^4$~T), $\bm{E}_\perp$ is suppressed and much smaller than the beam self-field ($9.4\times10^{12}$~V/m). 

Figure~\ref{fig:2}(d) [\figref{fig:2}(e)] plots $\eta$ during single electron beam-single foil collision with the same parameters as above but for $\sigma_\parallel=0.55\,\mu$m ($\sigma_\perp=0.55\,\mu$m ) and $\sigma_\perp$ ($\sigma_\parallel$) ranging from $0.275\,\mu$m to $1.25\,\mu$m. Black circles and blue circles correspond to 3D PIC simulation and reflected-field model results, respectively. These simulations confirm that the mechanism of beam focusing and photon emission is robust and effective. Indeed, as shown in the Supplemental Material, dense collimated photon beams can already be generated with the electron beam parameters attainable at existing accelerator facilities such as FACET-II~\cite{yakimenkoPRAB19}.

Figures~\ref{fig:2}(d)-(e) show that simulation results approach the prediction of the reflected-field model with increasing (decreasing) $\sigma_\perp$ ($\sigma_\parallel$). For beam density smaller than the foil electron density, simulations indicate that foil thickness is irrelevant provided that collisions and plasma instabilities remain negligible. By contrast, foil thickness is important when the electron beam density exceeds the conductor density~\cite{sampathThes}. Note that synchrotron photon emission also occurs when the beam exits the foil, as $E_r$ is suppressed at the rear foil surface and $B_\theta$ grows gradually during the beam exit~\cite{carronPIER00}. However, for $\sigma_\perp \gtrsim \sigma_\parallel$, the contribution of the rear surface to the radiated energy is subdominant, and is neglected in our model. 

\begin{figure}[tb]
\centering
\includegraphics[width=\linewidth]{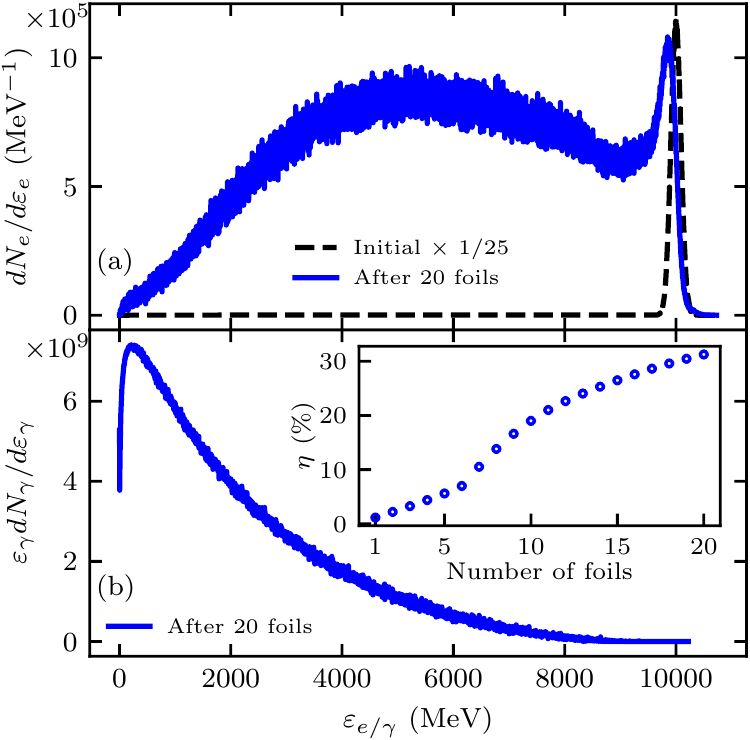} 
\caption{(a)~Initial (black dashed line) and final (blue line) electron beam energy distribution. (b)~Final photon spectrum. The inset displays $\eta$ as a function of the number of foils crossed by the electron beam.}
\label{fig:4}
\end{figure}

The above considerations suggest that the focusing and radiative effects can be substantially enhanced by colliding the self-focused beam with further foils. In fact, the increased beam density results in stronger self-fields, thereby amplifying both self-focusing and photon emission at successive collisions (see Figs.~\ref{fig:3}-\ref{fig:4}). Note that, for efficient self-focusing, the distance between two consecutive foils needs to be sufficiently large to allow beam self-field restoration around its propagation axis ($r \lesssim\sigma_\perp$). This requires that the travelled distance is much larger than $\sigma_\perp$. Furthermore, the interfoil distance needs to be short enough to prevent beam expansion. This can be estimated by considering the effect of $|\bm{f}_\perp| \approx |2 q B_\theta|$ calculated at $x \approx x_0 + v t$ and $r \approx \sigma_\perp$, i.e., where focusing is stronger. For $\sigma_\perp \gtrsim \sigma_\parallel$, CTR extends approximately over a distance $\sigma_\perp^2/\sigma_\parallel$, which is larger than the beam length $\sigma_\parallel$. Thus, $|\bm{f}_\perp|$ lasts for approximately $\sigma_\parallel/c$, and the deflection angle is $\vartheta \approx |\bm{f}_\perp| \sigma_\parallel/\gamma m_e c^2$. Hence, to prevent defocusing the interfoil distance must be much smaller than $\sigma_\perp/\vartheta$. Also, for effective focusing, $\vartheta$ must be much larger than the beam angular spread $\Delta \vartheta_{p_\perp/p_\parallel} \approx \epsilon_n/\gamma \sigma_\perp$, where $\epsilon_n$ is the normalized beam emittance.

In our multifoil 3D PIC simulations, the electron beam has 2~nC charge, Gaussian spatial and momentum distributions with $ \sigma_\parallel=\sigma_\perp=0.55\text{ $\mu$m}$, 10~GeV mean energy,  212~MeV FWHM energy spread, and 3~mm-mrad normalized emittance ($\vartheta \approx 2$~mrad with these parameters). Comparable parameters are expected at the advanced stage of FACET-II by employing a plasma lens \cite{yakimenkoPRAB19, dossPRAB19}. The beam collides with 20 consecutive aluminum foils with $0.5\,\mu$m thickness, $10\,\mu$m interfoil distance, and $1.8\times10^{29}\,\text{m}^{-3}$ initial electron density. The computational box size is $6.6\,\mu\text{m}(x) \times 8.8\,\mu\text{m}(y) \times 8.8\,\mu\text{m}(z)$ with $528(x) \times 352(y) \times 352(z)$ gridpoints, 4 particles-per-cell (ppc) for beam electrons and 8~ppc for foil electrons and ions were used. The moving window technique was employed to follow the beam evolution. Simulations were independently performed with Smilei~\cite{derouillatCPC18, smileiURL} and CALDER~\cite{lefebvreNF03} PIC codes with good agreement. The initial self-consistent beam fields, the effect of field and collisional ionization and binary Coulomb collisions were included. Synchrotron and bremsstrahlung emission, and multiphoton Breit-Wheeler and Bethe-Heitler pair production were implemented with state-of-the-art Monte-Carlo methods~\cite{sampathThes, smileiURL, lobetJPCS16, martinezPoP19}. Consistent with the submicrometer foil thickness, simulations showed that collisional processes are negligible. 

Figure~\ref{fig:3} displays snapshots of the electron and gamma-ray beam evolution (see Supplemental Material for a movie). 
Until the 6th foil, the beam interacts with the field ``reflected'' by each foil. This leads the beam to self-focus and gradually become denser (see the first to third column of \figref{fig:3}). The electron beam density rises from its initial value of $4.7\times10^{27}$~m$^{-3}$ to $8.2\times10^{28}$~m$^{-3}$ after the 6th foil, while the maximum photon density and $\chi$ are $2.9\times10^{28}$~m$^{-3}$ and 0.8, respectively [see \figref{fig:3}(a3)-(d3)]. During (immediately after) the interaction with the 7th foil, the electron beam density reaches $3.8\times10^{29}$~m$^{-3}$ ($4.5\times10^{29}$~m$^{-3}$), which exceeds the foil density of $1.8\times10^{29}$~m$^{-3}$. Hence, the foil is unable to reflect the fields of the beam, and a channel where foil electrons are expelled is created [see \figref{fig:3}(b4)-(c4)]. Here $\chi$ and the photon beam density rise up to 3 and $4.1\times10^{29}$~m$^{-3}$, respectively. The overall maximum gamma-ray density of $8.5\times10^{29}$~m$^{-3}$ is reached immediately after electron beam interaction with the 8th foil. Moreover, a fraction of approximately $10^{-4}$ photons with energies $>2 m_e c^2$ convert into $e^-e^+$ pairs via the multiphoton Breit-Wheeler process. Electron beam density stops increasing when it becomes larger than the foil electron density. In the following beam-foil collisions, the electron beam density profile undergoes longitudinal modulations, the reflected-field strength being dependent on the longitudinal position and stronger around the rear parts of the beam [\figref{fig:3}(a5)-(a6)].

Figure~\ref{fig:4}(a) plots the initial (black dashed line) and final (blue line) electron beam energy distribution after the interaction with 20 consecutive foils. The broad distribution around approximately 5~GeV results from intense synchrotron emission occurring in the central and rear part of the electron beam. The residual peak around the initial electron beam energy is indicative of the small synchrotron and collisional energy losses in the front part of the beam, which experiences only weak amplitude CTR. Figure~\ref{fig:4}(b) reports the final photon spectrum and the conversion efficiency $\eta$ (inset) as a function of the number of crossed foils. The increase in $\eta$ at the 7th foil is due to the extremely high beam density and, consequently, to the ultrastrong fields induced inside the foil. After colliding with 20 foils, more than 30\% of the electron beam energy is converted into a collimated (5~mrad rms photon energy angular distribution), 4~fs FWHM duration, $2.8\times10^{29}$~m$^{-3}$ peak density gamma-ray pulse.

In summary, we have introduced a new scheme to efficiently produce extremely dense gamma-ray beams from the interaction of a high-current ultrarelativistic electron beam with a sequence of thin foils. This scheme also provides a promising route for producing solid-density ultrarelativistic electron beams and for exploring strong-field QED processes with a single electron beam, that is, without the need of an external powerful laser drive. In fact, following a methodology analogous to that employed with intense laser pulses~\cite{yanNP17, colePRX18, poderPRX18}, the field experienced \textit{in situ} by the electron beam and the ensuing strong-field QED effects can be inferred by measuring particle angular distributions, spectra, and photon and pair yields along with CTR~\cite{vantilborgPRL06, glinecPRL07, maxwellPRL13, lundhPRL13}.

\begin{acknowledgments}
This work was performed in the framework of the E-305 collaboration. E-305 is a SLAC experiment whose aims include the generation of bright gamma rays, in particular in electron beam-solid interaction. Based on the findings of this work, the E-332 experiment on solid-density gamma-ray pulse generation in electron beam-multifoil interaction has been developed and approved, and will be carried out at SLAC. This article comprises parts of the Ph.D. thesis work of Archana Sampath, submitted to the Heidelberg University, Germany.

The work at LOA was supported by the European Research Council (ERC) under the European Union’s Horizon 2020 research and innovation programme (Grant Agreement No. 715807). We acknowledge GENCI for granting us access to the supercomputer Irene under the grant No. A0080510786 to run CALDER simulations. The work at SLAC was supported by U.S. DOE FES grant No. FWP100331. UCLA was supported by U.S. Department of Energy grant  No. DE-SC001006 and NSF grant  No. 1734315.
\end{acknowledgments}

\bibliography{My_Bibliography}

\clearpage


\section*{Supplementary material}

# Supplemental Material for
# "Extremely Dense Gamma-Ray Pulses in Electron Beam-Multifoil Collisions"

Archana Sampath,[1] Xavier Davoine,[2,3] Sébastien Corde,[4] Laurent Gremillet,[2,3] Max Gilljohann,[4] Maitreyi Sangal,[1] Christoph H. Keitel,[1] Robert Ariniello,[5] John Cary,[5] Henrik Ekerfelt,[6] Claudio Emma,[6] Frederico Fiuza,[6] Hiroki Fujii,[7] Mark Hogan,[6] Chan Joshi,[7] Alexander Knetsch,[4] Olena Kononenko,[4] Valentina Lee,[5] Mike Litos,[5] Kenneth Marsh,[7] Zan Nie,[7] Brendan O'Shea,[6] J. Ryan Peterson,[6,8] Pablo San Miguel Claveria,[4] Doug Storey,[6] Yipeng Wu,[7] Xinlu Xu,[6] Chaojie Zhang,[7] and Matteo Tamburini[1, *]

[1]*Max-Planck-Institut für Kernphysik, Saupfercheckweg 1, D-69117 Heidelberg, Germany*
[2]*CEA, DAM, DIF, 91297 Arpajon, France*
[3]*Université Paris-Saclay, CEA, LMCE, 91680 Bruyères-le-Châtel, France*
[4]*LOA, ENSTA Paris, CNRS, Ecole Polytechnique,*
*Institut Polytechnique de Paris, 91762 Palaiseau, France*
[5]*University of Colorado Boulder, Department of Physics,*
*Center for Integrated Plasma Studies, Boulder, Colorado 80309, USA*
[6]*SLAC National Accelerator Laboratory, Menlo Park, CA 94025, USA*
[7]*University of California Los Angeles, Los Angeles, CA 90095, USA*
[8]*Stanford University, Physics Department, Stanford, CA 94305, USA*
(Dated: December 18, 2020)

## S1. TRANSITION RADIATION BY AN ULTRARELATIVISTIC PARTICLE BEAM IN THE NEAR-FIELD ZONE

### A. Pseudophoton method

Let us consider an ultrarelativistic particle beam with cylindrical symmetry around its propagation axis and moving along $x > 0$ at a velocity $v$ in the laboratory frame. In its rest frame, the beam density distribution is chosen as

$$n'(\mathbf{x}') = n_0' e^{-r'^2/2\sigma_\perp'^2 - x'^2/2\sigma_\parallel'^2} . \tag{S1}$$

Here primed quantities refer to their values in the beam rest frame, $n_0' = N_e/(2\pi)^{3/2}\sigma_\perp'^2\sigma_\parallel'$, and $N_e$ is the number of beam particles of individual charge $q$. We assume that the longitudinal and transverse (rms) beam dimensions fulfill $\sigma_\parallel' = \gamma\sigma_\parallel \gg \sigma_\perp' = \sigma_\perp$, where $\gamma = (1 - v^2/c^2)^{-1/2}$ is the beam relativistic factor. Hence, the beam self-electric field can be evaluated from Gauss' law in the long-beam limit

$$\mathbf{E}'^b(\mathbf{x}') \approx \mathbf{E}_\perp^b(\mathbf{x}') \approx 4\pi n_0' q\sigma_\perp'^2 \frac{\left(1 - e^{-r'^2/2\sigma_\perp'^2}\right)}{r} e^{-x'^2/2\sigma_\parallel'^2} \hat{\mathbf{r}} , \tag{S2}$$

which is valid for $r' \ll \sigma_\parallel'$, i.e., $r \ll \gamma\sigma_\parallel$. An inverse Lorentz transformation yields the beam self-field in the laboratory frame

$$\mathbf{E}_\perp^b(x, r, t) \approx 4\pi n_0 q\sigma_\perp^2 \frac{\left(1 - e^{-r^2/2\sigma_\perp^2}\right)}{r} e^{-(x-vt)^2/2\sigma_\parallel^2} \hat{\mathbf{r}} . \tag{S3}$$

Here we have introduced the beam density in the laboratory $n_0 = \gamma n_0'$ and, for simplicity, the time origin is chosen such that at $t = 0$ the beam center is located at $x = 0$.

By using the Weizsäcker-Williams approximation [1] the beam electric self-field can be decomposed into plane waves (pseudophotons) whose electric components are

$$\widetilde{\mathbf{E}}_\perp^b(x, \mathbf{k}_\perp, \omega) = \iint d^2\mathbf{r}_\perp dt \, \mathbf{E}_\perp^b(x, r, t) e^{-i\mathbf{k}_\perp \cdot \mathbf{r}_\perp + i\omega t} = -8i\sqrt{2}\pi^{5/2} n_0 q \frac{\sigma_\parallel \sigma_\perp^2}{k_\perp v} e^{i\omega x/v} e^{-\sigma_\parallel^2\omega^2/2v^2 - \sigma_\perp^2 k_\perp^2/2} \hat{\mathbf{k}}_\perp , \tag{S4}$$

---
* matteo.tamburini@mpi-hd.mpg.de

where we have used the integral formulae [2]

$$\int_0^{2\pi} d\phi \, \cos\phi e^{ik_\perp r \cos\phi} = 2i\pi \, J_1(k_\perp r) \,, \tag{S5}$$

$$\int_0^\infty dr \, J_\nu(k_\perp r) = 1/k_\perp \,, \tag{S6}$$

$$\int_0^\infty dr e^{-r^2/2\sigma_\perp^2} \, J_1(k_\perp r) = \sqrt{\frac{\pi}{2}} \sigma_\perp e^{-\sigma_\perp^2 k_\perp^2/4} I_{1/2} \left(\sigma_\perp^2 k_\perp^2/4\right) \,, \tag{S7}$$

with $I_{1/2}(z) = \sqrt{2/\pi} \sinh(z)/\sqrt{z}$.

In the case of a perfect conductor with front surface at $x = 0$, the sum of the transverse incident ($\widetilde{\mathbf{E}}_\perp^b$) and induced/scattered conductor ($\widetilde{\mathbf{E}}_\perp^c$) electric fields should vanish at the boundary, so that

$$\widetilde{\mathbf{E}}_\perp^c(0^-, \mathbf{k}_\perp, \omega) = -\widetilde{\mathbf{E}}_\perp^b(0^-, \mathbf{k}_\perp, \omega) \,. \tag{S8}$$

The scattered field can then be exactly computed at any time and position in vacuum ($x < 0$) by using the plane-wave decomposition [3]

$$\begin{aligned}
\mathbf{E}_\perp^c(x, \mathbf{r}_\perp, t) &= (2\pi)^{-3} \iint d\omega d^2\mathbf{k}_\perp \, \widetilde{\mathbf{E}}_\perp^c(0^-, \mathbf{k}_\perp, \omega) \, e^{-i\omega t + ik_x x + i\mathbf{k}_\perp \cdot \mathbf{r}_\perp} \\
&= i\sqrt{\frac{2}{\pi}} \frac{n_0 q}{v} \sigma_\parallel \sigma_\perp^2 \int_{-\infty}^\infty d\omega \, e^{-\sigma_\parallel^2 \omega^2/2v^2 - i\omega t} \int \frac{d^2\mathbf{k}_\perp}{k_\perp} \, e^{-\sigma_\perp^2 k_\perp^2/2 + i\mathbf{k}_\perp \cdot \mathbf{r}_\perp + ik_x x} \, \widehat{\mathbf{k}}_\perp \,.
\end{aligned} \tag{S9}$$

We have introduced the longitudinal wavenumber

$$k_x = \begin{cases} -\mathrm{sgn}(\omega)\sqrt{\omega^2/c^2 - k_\perp^2} & \text{if } \omega^2/c^2 > k_\perp^2 \\ -i\sqrt{k_\perp^2 - \omega^2/c^2} & \text{if } \omega^2/c^2 < k_\perp^2 \end{cases} \tag{S10}$$

In Eq. (S9), Fourier components with $\omega^2/c^2 > k_\perp^2$ describe propagating waves, while components with $\omega^2/c^2 < k_\perp^2$ correspond to evanescent waves confined to the conductor's boundary [4]. The choice of the minus sign in the definition of $k_x$ is consistent with waves propagating/damped along $x < 0$.

Equation (S9) can be simplified by expressing the transverse wavenumber as $\mathbf{k}_\perp = k_\perp(\cos\phi \, \widehat{\mathbf{r}} + \sin\phi \, \widehat{\boldsymbol{\phi}})$ and performing the integration over $\phi$, which gives

$$\mathbf{E}_\perp^c(x, r, t) = E_r^c(x, r, t)\widehat{\mathbf{r}} = -\sqrt{8\pi} \frac{n_0 q}{v} \sigma_\parallel \sigma_\perp^2 \int_{-\infty}^\infty d\omega \, e^{-\sigma_\parallel^2 \omega^2/2v^2 - i\omega t} \int_0^\infty dk_\perp \, e^{-\sigma_\perp^2 k_\perp^2/2 + ik_x x} J_1(k_\perp r)\widehat{\mathbf{r}} \,. \tag{S11}$$

This equation is accurate in the domain of validity of the field approximation in Eq. (S3), and therefore holds in the near beam region that is of central interest here.

The longitudinal electric field follows from Gauss' law in vacuum [5],

$$\frac{\partial E_x^c}{\partial x} = -\frac{1}{r} \frac{\partial}{\partial r} \left(r E_r^c\right) \,, \tag{S12}$$

yielding

$$E_x^c(x, r, t) = -i\sqrt{8\pi} \frac{n_0 q}{v} \sigma_\parallel \sigma_\perp^2 \int_{-\infty}^\infty d\omega \, e^{-\sigma_\parallel^2 \omega^2/2v^2 - i\omega t} \int_0^\infty dk_\perp \frac{k_\perp}{k_x} e^{-\sigma_\perp^2 k_\perp^2/2 + ik_x x} J_0(k_\perp r) \,. \tag{S13}$$

Finally, the induced magnetic field is given by Faraday's law

$$\frac{1}{c} \frac{\partial B_\theta^c}{\partial t} = \frac{\partial E_x^c}{\partial r} - \frac{\partial E_r^c}{\partial x} \,, \tag{S14}$$

and by using Eq. (S10) one obtains

$$B_\theta^c(x, r, t) = -\sqrt{8\pi} \frac{n_0 q}{vc} \sigma_\parallel \sigma_\perp^2 \int_{-\infty}^\infty d\omega \, \omega e^{-\sigma_\parallel^2 \omega^2/2v^2 - i\omega t} \int_0^\infty \frac{dk_\perp}{k_x} e^{-\sigma_\perp^2 k_\perp^2/2 + ik_x x} J_1(k_\perp r) \,. \tag{S15}$$

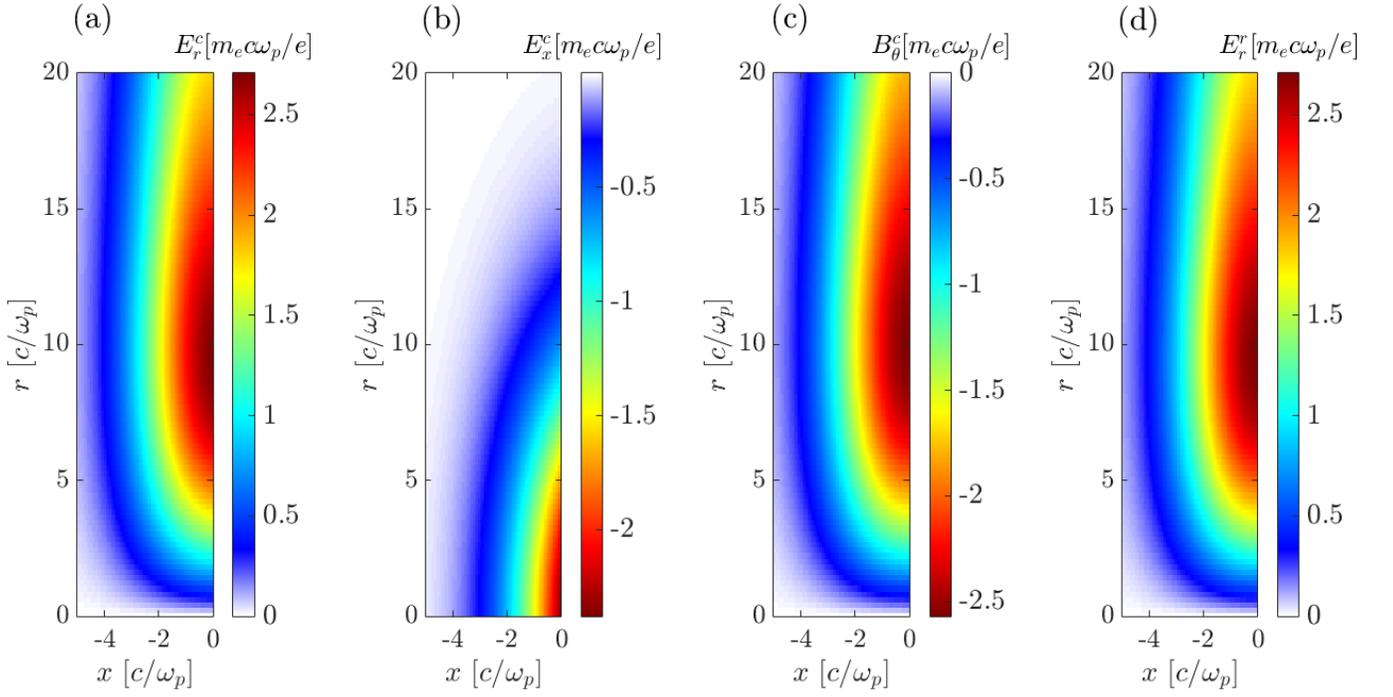

FIG. S1. (a) Transverse and (b) longitudinal electric and (c) magnetic field of the conductor, compared with (d) transverse electric field of the beam's image charge at $\omega_p t = 0$. Beam parameters are $\gamma = 2 \times 10^4$, $\sigma_\parallel \omega_p / c = 2$ and $\sigma_\perp \omega_p / c = 6$.

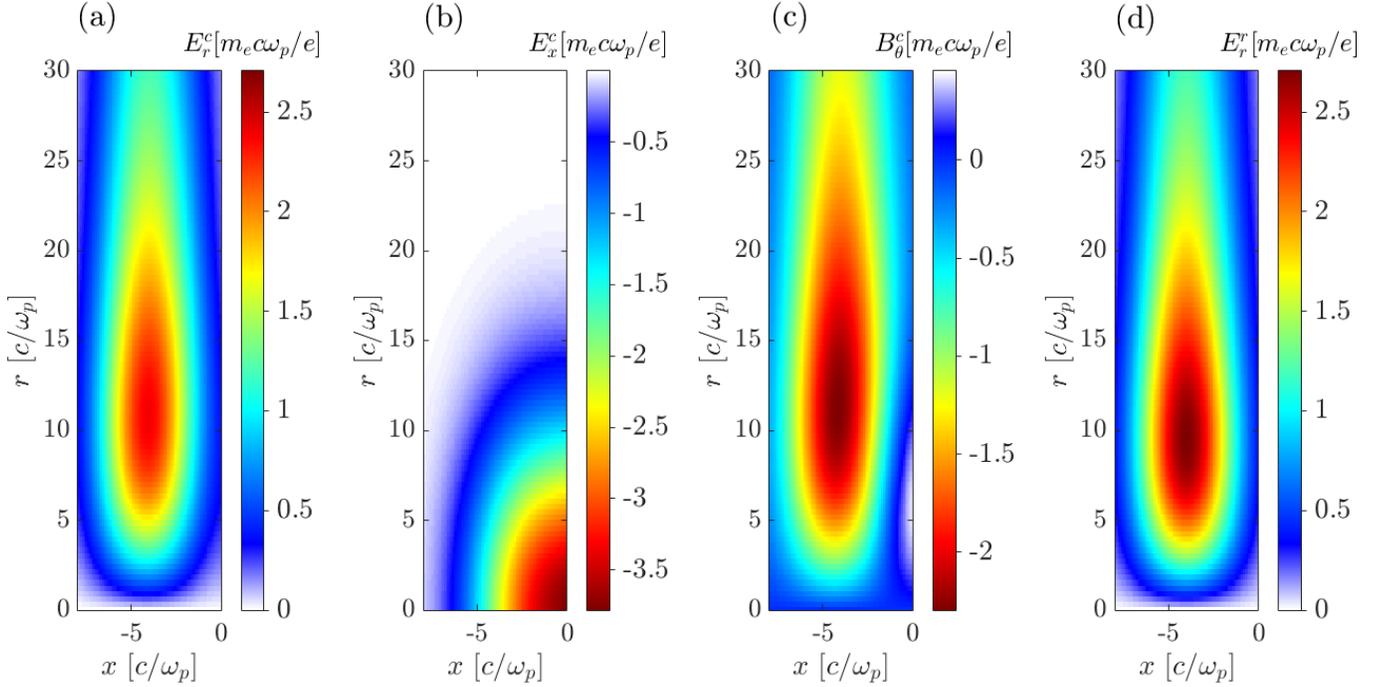

FIG. S2. Same as Fig. S1 but at a later time, $\omega_p t = 4$.

Figures S1 and S2 display 2D maps of the conductor electric (a)-(b) and magnetic (c) fields as computed numerically from Eqs. (S11), (S13), and (S15). We first consider an ultrarelativistic electron ($q = -e$) beam with $\gamma = 2 \times 10^4$, $\sigma_\parallel \omega_p / c = 2$ and $\sigma_\perp \omega_p / c = 6$, where $\omega_p = \sqrt{4\pi n_0 e^2 / m_e}$ is the beam plasma frequency. These fields are compared with the radial electric field of the beam's image charge $E_r^r$ (d) (the "reflected field"), obtained by setting $q \to -q$ and $v \to -v$ in Eq. (S3). At $t = 0$ (see Fig. S1), both the transverse conductor electric and magnetic fields $E_r^c$

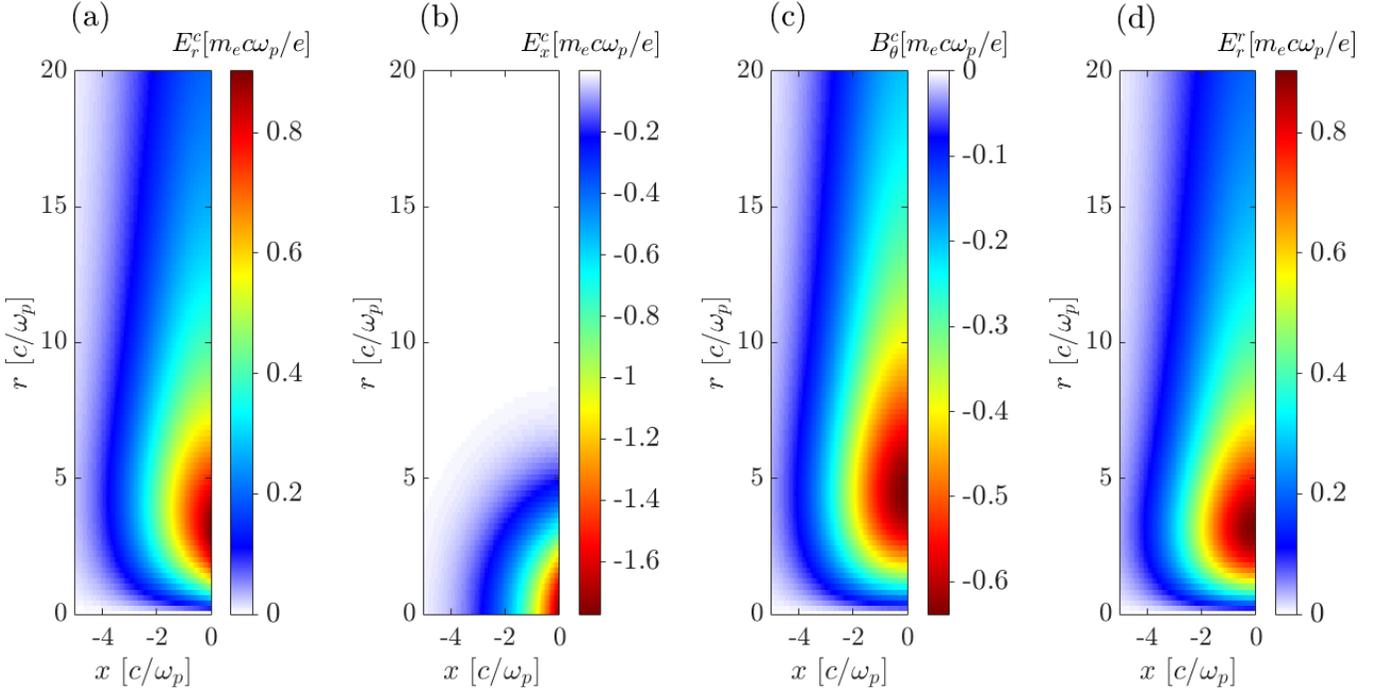

FIG. S3. (a) Transverse and (b) longitudinal electric and (c) magnetic field of the conductor, compared with (d) transverse electric field of the beam's image charge at $\omega_p t = 0$. Beam parameters are $\gamma = 2 \times 10^4$, $\sigma_\parallel \omega_p/c = 2$ and $\sigma_\perp \omega_p/c = 2$.

and $B^c_\theta$ are very similar to the reflected field around the boundary ($E^c_r \approx -B^c_\theta \approx E^r_r \approx 2.7\,m_e c \omega_p/e$). The induced longitudinal electric field has a maximum amplitude slightly lower ($E^c_x \approx -2.4\,m_e c \omega_p/e$), reached at $r = 0$ (in contrast to the transverse fields which peak at $r \approx 1.6\,\sigma_\perp \approx 9.5\,c/\omega_p$). At $\omega_p t = 4$ (see Fig. S2), the transverse induced fields detach from the boundary as is consistent for a propagating wavepacket. Yet diffraction causes these radiated fields to drop in amplitude ($E^c_r \approx 2.4\,m_e c \omega_p/e$ and $B^c_\theta \approx -2.3\,m_e c \omega_p/e$), and to reach their maxima at a larger radius ($r \omega_p/c \approx 11$) than the reflected field. The longitudinal induced field remains mainly confined to the boundary and has been intensified ($E^c_x \approx -3.8\,m_e c \omega_p/e$). Around the boundary and $r \approx \sigma_\perp$, the magnetic field is then mainly driven through $\partial B^c_\theta/c\partial t \approx \partial E^c_x/\partial r > 0$, and so becomes of positive polarity [see Fig. S2(c)]

Figures S3(a)-(d) plot the same quantities as Figs. S1-S2 but for a spherical beam ($\sigma_\parallel = \sigma_\perp = 2c/\omega_p$) and at $t = 0$. The scattered fields show a substantial divergence because the dominant wavenumber $\lambda \sim \sigma_\parallel$ is close to the transverse beam size $\sigma_\perp$. This translates into a longitudinal electric field $E^c_x$ about twice stronger than $E^c_r$. The amplitude of the scattered magnetic field is close to, yet a bit lower than the beam self-field ($B^c_\theta \approx -0.64\,m_e c \omega_p/e$ vs. $B^b_\theta \approx 0.90\,m_e c \omega_p/e$).

The case of a longitudinally elongated beam ($\sigma_\parallel \omega_p/c = 6$, $\sigma_\perp \omega_p/c = 1$) is illustrated at two successive times in Figs. S4 and S5. The small $\sigma_\perp/\sigma_\parallel$ ratio leads to most of the induced fields being evanescent, and thus strongly localized at the conductor's boundary. At both $t = 0$ (see Fig. S4) and $\omega_p t = 5$ (see Fig. S5), the longitudinal electric field exceeds its transverse counterpart. In parallel, the induced magnetic field gets considerably weaker than the magnetic self-field of the beam, and changes polarity along the boundary [see Fig. S5(c)].

## B. Evaluation of the surface magnetic field in the radiating and nonradiating regimes

We now demonstrate that the magnetic field induced at the boundary equals the beam self-field when $\sigma_\perp \gg \sigma_\parallel$ and tends to vanish when $\sigma_\perp \ll \sigma_\parallel$. To this purpose, we start by the expression of $B^c_\theta$ at $x = 0^-$

$$B^c_\theta(0^-, r, t) = -\sqrt{8\pi}\frac{n_0 q}{vc}\sigma_\parallel \sigma_\perp^2 \int_{-\infty}^{\infty} d\omega\, \omega e^{-\sigma_\parallel^2 \omega^2/2v^2 - i\omega t} \int_0^{\infty} \frac{dk_\perp}{k_x} e^{-\sigma_\perp^2 k_\perp^2/2} J_1(k_\perp r)\,. \tag{S16}$$

When $\sigma_\perp \gg \sigma_\parallel$, the dominant integration range over $k_\perp$ is such that $k_\perp^2 \ll \omega^2/c^2 \approx 2/\sigma_\parallel^2$, so that the scattered

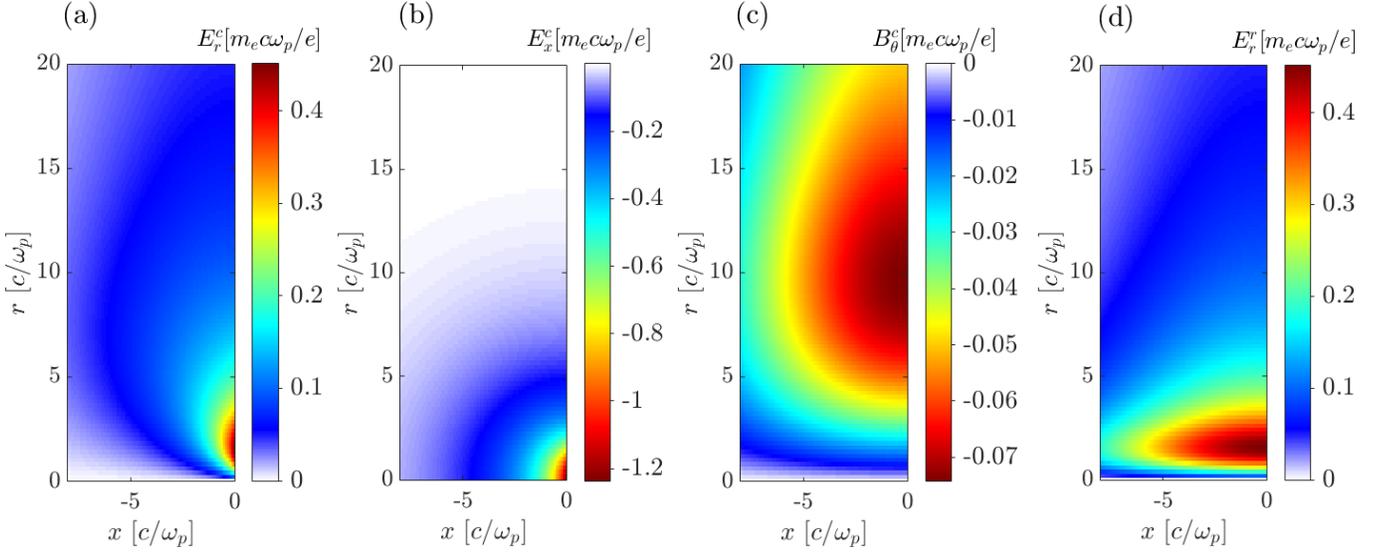

FIG. S4. (a) Transverse and (b) longitudinal electric and (c) magnetic field of the conductor, compared with (d) transverse electric field of the beam's image charge at $\omega_p t = 0$. Beam parameters are $\gamma = 2 \times 10^4$, $\sigma_\parallel \omega_p/c = 6$ and $\sigma_\perp \omega_p/c = 1$.

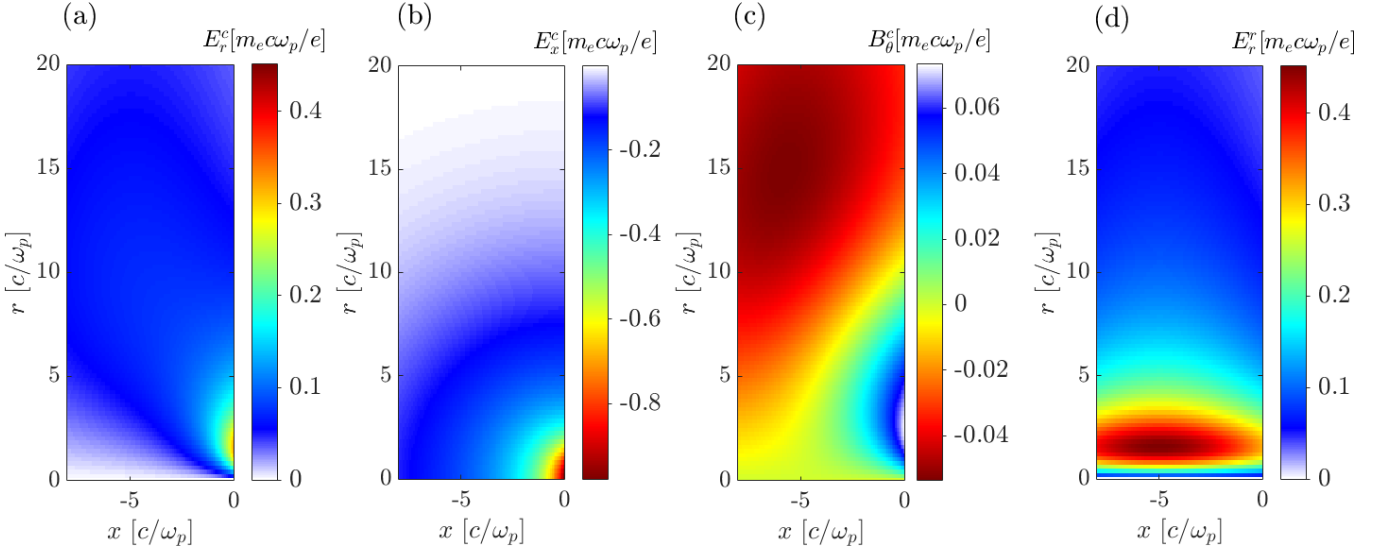

FIG. S5. Same as Fig. S4 but at a later time, $\omega_p t = 5$.

field mainly consists of propagating waves. Approximating $k_x \approx -\omega/c$ yields

$$B_\theta^c(0^-, r, t) \approx \sqrt{8\pi} \frac{n_0 q}{v} \sigma_\parallel \sigma_\perp^2 \int_{-\infty}^{\infty} d\omega \, e^{-\sigma_\parallel^2 \omega^2 / 2v^2 - i\omega t} \int_0^{\infty} dk_\perp \, e^{-\sigma_\perp^2 k_\perp^2 / 2} J_1(k_\perp r) \,. \tag{S17}$$

It is easily seen that the above approximate expression of $B_\theta^c(0^-, r, t)$ exactly coincides with $-E_r^c(0^-, r, t)$ from Eq. (S11). Since by construction $E_r^c(0^-, r, t) = -E_r^b(0^-, r, t) = -B_\theta^b(0^-, r, t)$, we conclude that $B_\theta^c(0^-, r, t) = B_\theta^b(0^-, r, t)$ in the radiating regime.

The calculation is more involved when $\sigma_\perp \ll \sigma_\parallel$. By using Eq. (S10), we first decompose the integral over $k_\perp$ in Eq. (S16) as

$$\int_0^{\infty} \frac{dk_\perp}{k_x} e^{-\sigma_\perp^2 k_\perp^2 / 2} J_1(k_\perp r) = -\mathrm{sgn}(\omega) \int_0^{|\omega|/c} dk_\perp \frac{J_1(k_\perp r)}{\sqrt{\omega^2/c^2 - k_\perp^2}} e^{-\sigma_\perp^2 k_\perp^2 / 2}$$

$$+ i \int_{|\omega|/c}^{\infty} dk_\perp \frac{J_1(k_\perp r)}{\sqrt{k_\perp^2 - \omega^2/c^2}} e^{-\sigma_\perp^2 k_\perp^2 / 2} \equiv \mathcal{I}_1 + \mathcal{I}_2 \,. \tag{S18}$$

where the integrals $\mathcal{I}_1$ and $\mathcal{I}_2$ account for the propagating and non-propagating (evanescent) modes, respectively. In $\mathcal{I}_1$, one can neglect the variation of the exponential term because $\sigma_\perp k_\perp \leq \sigma_\perp \omega/c \sim (\sigma_\perp/\sigma_x) \ll 1$, and thus approximate this integral as

$$\mathcal{I}_1 \approx -\text{sgn}(\omega) \int_0^{|\omega|/c} dk_\perp \frac{J_1(k_\perp r)}{\sqrt{\omega^2/c^2 - k_\perp^2}} = -\text{sgn}(\omega) \int_0^1 du \frac{J_1(|\omega|ru/c)}{\sqrt{1-u^2}} = -\frac{\sin^2(\omega r/2c)}{(\omega r/2c)} \,. \tag{S19}$$

The latter equality follows from Eq. 6.552-7 in [2], and by noting that $\sin^2(|\omega|r/2c) = \sin^2(\omega r/2c)$, and $\omega = \text{sgn}(\omega)|\omega|$. To evaluate $\mathcal{I}_2$, we first recast it as

$$\mathcal{I}_2 = i \int_1^\infty du \frac{J_1(|\omega|ru/c)}{\sqrt{u^2-1}} e^{-\sigma_\perp^2 \omega^2 u^2/2c^2} \,. \tag{S20}$$

Let us first consider radial positions $r \ll \sigma_\perp$. In this case, the effective integration range is $[1, \sim \sqrt{2}c/\sigma_\perp|\omega|]$, and one can expand the Bessel function around zero $J_1(x) \approx x/2 + \mathcal{O}(x^3)$. This yields

$$\mathcal{I}_2 \approx i\frac{|\omega|r}{2c} \int_1^\infty du \frac{u\, e^{-(\sigma_\perp \omega u/c)^2/2}}{\sqrt{u^2-1}} = i\frac{|\omega|r}{4c} \int_1^\infty dw \frac{e^{-(\sigma_\perp \omega/c)^2 w/2}}{\sqrt{w-1}} = i\sqrt{\frac{\pi}{8}} \frac{r}{\sigma_\perp} e^{-(\sigma_\perp \omega/c)^2/2} \,. \tag{S21}$$

By plugging Eqs. (S19) and (S21) into Eq. (S16) and using $e^{-i\omega t} = \cos(\omega t) + (i/\omega)d\cos(\omega t)/dt$ one obtains

$$\begin{aligned}
B_\theta^c(0^-, r \ll \sigma_\perp, t) \approx \sqrt{8\pi} \frac{n_0 q}{vc} \sigma_\parallel \sigma_\perp^2 &\left[ \frac{2c}{r} \int_{-\infty}^\infty d\omega\, e^{-(\sigma_\parallel \omega/v)^2/2} \sin^2(\omega r/2c) \cos(\omega t) \right.\\
&\left. + \sqrt{\frac{\pi}{8}} \frac{r}{\sigma_\perp} \frac{d}{dt} \int_{-\infty}^\infty d\omega\, e^{-(\sigma_\parallel^2/v^2 + \sigma_\perp^2/c^2)\omega^2/2} \cos(\omega t) \right] \,.
\end{aligned} \tag{S22}$$

After some calculations, one obtains

$$\begin{aligned}
B_\theta^c(0^-, r \ll \sigma_\perp, t) \approx 2\pi n_0 q r &\left[ \left(\frac{\sigma_\perp}{r}\right)^2 \left( 2e^{-(vt/\sigma_\parallel)^2/2} - e^{-(v/\sigma_\parallel)^2(t-r/c)^2/2} - e^{-(v/\sigma_\parallel)^2(t+r/c)^2/2} \right) \right.\\
&\left. - \sqrt{\frac{\pi}{2}} \frac{\sigma_\parallel \sigma_\perp t}{vc(\sigma_\parallel^2/v^2 + \sigma_\perp^2/c^2)^{3/2}} e^{-t^2/2(\sigma_\parallel^2/v^2 + \sigma_\perp^2/c^2)} \right] \,.
\end{aligned} \tag{S23}$$

Since $v \approx c$ and $\sigma_\parallel \gg \sigma_\perp$, the above equation can be further simplified as

$$\begin{aligned}
B_\theta^c(0^-, r \ll \sigma_\perp, t) \approx 2\pi n_0 q r &\left[ \left(\frac{\sigma_\perp}{r}\right)^2 \left( 2e^{-(ct/\sigma_\parallel)^2/2} - e^{-(c/\sigma_\parallel)^2(t-r/c)^2/2} - e^{-(c/\sigma_\parallel)^2(t+r/c)^2/2} \right) \right.\\
&\left. - \sqrt{\frac{\pi}{2}} \frac{\sigma_\perp ct}{\sigma_\parallel^2} e^{-(ct/\sigma_\parallel)^2/2} \right] \,.
\end{aligned} \tag{S24}$$

From Eq. (S24) and expanding for small $r$, it follows that when the central beam slice reaches the conductor $(t = 0)$, the induced magnetic field close to the axis has an amplitude

$$B_\theta^c(0^-, r \ll \sigma_\perp, 0) \approx 2\pi n_0 q r \left(\frac{\sigma_\perp}{\sigma_\parallel}\right)^2 + \mathcal{O}(r^3) \approx \left(\frac{\sigma_\perp}{\sigma_\parallel}\right)^2 B_\theta^b(0^-, r \ll \sigma_\perp, 0) \,, \tag{S25}$$

where in the last equality we used the small $r$ expansion of Eq. (S3) and $\mathbf{B}^b = \mathbf{v}/c \times \mathbf{E}^b$. The surface magnetic field of the conductor at $t = 0$ is therefore $(\sigma_\parallel/\sigma_\perp)^2$ lower than the beam self-field. The accuracy of this approximation can be assessed for the parameters of Fig. S4. For $\omega_p \sigma_\parallel/c = 6$, $\omega_p \sigma_\perp/c = 1$, $\omega_p r/c = 0.1$ and $\omega_p t = 0$, the numerical evaluation of Eq. (S16) gives $B_\theta^c \approx 1.35 \times 10^{-3} m_e \omega_p/e$, close to the above formula, $B_\theta^c \approx 1.39 \times 10^{-3} m_e \omega_p/e$.

By expanding Eq. (S24) for small $r$ one gets

$$B_\theta^c(0^-, r \ll \sigma_\perp, t) \approx -2\pi n_0 q r e^{-c^2 t^2/2\sigma_\parallel^2} \left( \frac{\sigma_\perp^2}{\sigma_\parallel^2} \frac{c^2 t^2}{\sigma_\parallel^2} + \sqrt{\frac{\pi}{2}} \frac{\sigma_\perp}{\sigma_\parallel} \frac{ct}{\sigma_\parallel} - \frac{\sigma_\perp^2}{\sigma_\parallel^2} \right) \,. \tag{S26}$$

Thus, when $t \gg r/c$ and $t \ll \sigma_\parallel^2/c\sigma_\perp$, which corresponds to the magnetostatic limit, the right-hand side of Eq. (S24) is dominated by the second term,

$$B_\theta^c(0^-, r \ll \sigma_\perp, t) \approx -\sqrt{2\pi^3} n_0 q \frac{\sigma_\perp r}{\sigma_\parallel^2} ct\, e^{-(ct/\sigma_\parallel)^2/2} \,, \tag{S27}$$

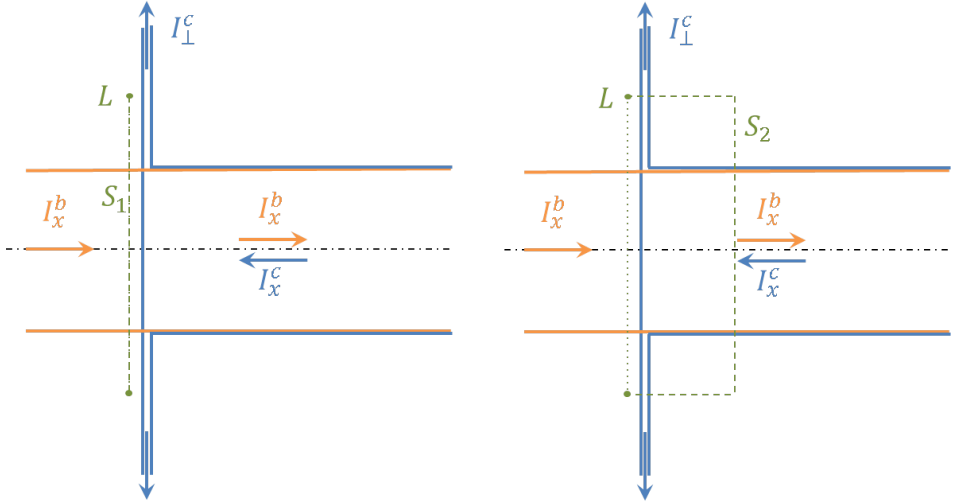

FIG. S6. Equilibrium current distributions due to a particle beam propagating into a conductor. Ampère's law is applied for two different surfaces $S_1$ and $S_2$ (dashed green curve) bounded by the loop $L$ (dotted green line).

that is, by the evanescent modes. The above expression could have been derived more directly by neglecting $\mathcal{I}_1$ in Eq. (S18) and approximating $k_x \sim -ik_\perp$ in $\mathcal{I}_2$.

Let us now estimate the surface magnetic field generated at large radii, $r \gg \sigma_\perp$. Equation Eq. (S20) can then be approximated as

$$\mathcal{I}_2 \approx i \int_1^\infty du \frac{J_1(|\omega|ru/c)}{\sqrt{u^2-1}} = i\frac{\sin(\omega r/c)}{\omega r/c} \ . \tag{S28}$$

By substituting Eq. (S28) and Eq. (S19) in Eq. (S16), and by using Euler's formula and the trigonometric identity $2\sin^2(\omega r/2c)\cos(\omega t) - \sin(\omega r/c)\sin(\omega t) = \cos(\omega t) - \cos(\omega(t-r/c))$, one obtains

$$B_\theta^c(0^-, r \gg \sigma_\perp, t) \approx -\sqrt{8\pi}n_0 q \frac{\sigma_\parallel \sigma_\perp^2}{vr} \int_{-\infty}^\infty d\omega \, e^{-(\sigma_\parallel \omega/v)^2/2} \left[\cos(\omega(t-r/c)) - \cos(\omega t)\right] \ . \tag{S29}$$

By explicitly calculating the integral one gets

$$B_\theta^c(0^-, r \gg \sigma_\perp, t) \approx 4\pi n_0 q \frac{\sigma_\perp^2}{r} \left[e^{-(vt/\sigma_\parallel)^2/2} - e^{-(v/\sigma_\parallel)^2(t-r/c)^2/2}\right] \ . \tag{S30}$$

From Eq. (S30) and since $v \approx c$, it follows that when the central beam slice reaches the conductor ($t = 0$), at large radii the magnetic field of the conductor scales as

$$B_\theta^c(0, r \gg \sigma_\perp, 0) \approx 4\pi n_0 q \frac{\sigma_\perp^2}{r} \left(1 - e^{-r^2/2\sigma_\parallel^2}\right) \ . \tag{S31}$$

This implies that for $\sigma_\perp \ll r \lesssim \sigma_\parallel$, $B^c(0^-, r, 0)$ is lower than the magnetic self-field of the beam by a factor of $[1 - \exp(-r^2/2\sigma_\parallel^2)]$, and that for $r \gg \sigma_\parallel \gg \sigma_\perp$, $B^c(0^-, r, 0)$ approaches the (radially decreasing) beam self-field. Interestingly, Eq. (S30) also indicates that $B_\theta^c(0^-, r \gg \sigma_\perp, t)$ changes its sign at $t = r/2c$, in agreement with Fig. S5.

Equations (S24) and (S30) demonstrate that the induced magnetic field vanishes everywhere along the conductor's surface when $\sigma_\parallel \to \infty$, that is, in the nonradiating (stationary) limit. This result can be readily understood from Fig. S6, which sketches the stationary current distribution induced in the conductor by the incoming particle beam. To ensure charge conservation, the longitudinal current induced through the beam cross-section deep inside the conductor must convert to a radial current along the conductor's boundary. Integrating the azimuthal magnetic field over a loop centered on $r = 0$ and tangential to the conductor's surface (see Fig. S6), Ampère's law tells us that

$$\int_L \mathbf{B} \cdot d\mathbf{l} = 2\pi r B_\theta(x = 0^-, r) = \frac{4\pi}{c} \int_S \mathbf{j} \cdot d\mathbf{S} \tag{S32}$$

where the surface $S$ is bounded by the loop $L$, and we have neglected the displacement current as we consider the stationary regime. By choosing $S$ as $S_1$ (see Fig. S6, left panel) gives

$$2\pi r B_\theta(x = 0^-, r) = \frac{4\pi}{c} \int_{S_1} \mathbf{j}_x^b dS = \frac{4\pi}{c} I_x^b. \tag{S33}$$

This means that $B_\theta(0^-, r)$ is only due to the beam current, and conversely that the induced currents in the conductor do not create a magnetic field outside of it. An alternative picture is provided by choosing the integration surface as $S_2$ (see Fig. S6, right panel)

$$2\pi r B_\theta(x = 0^-, r) = \frac{4\pi}{c} \left( I_x^b - I_x^c + I_\perp^c \right). \tag{S34}$$

In the steady state, one has $I_x^c = I_\perp^c$, which again implies a vanishing contribution of the conductor to $B_\theta(x = 0^-, r)$. We recall that Ampère's law in Eq. (S32) is valid only in the stationary case as it neglects the contribution of the displacement current. Thus, the reasoning of the stationary case does not apply to the radiating regime.

## S2. CONVERSION EFFICIENCY WITH LOWER DENSITY BEAMS

Simulations show that electron beams with much lower peak density, corresponding to more experimentally accessible conditions, can still generate radiation with electron-to-gamma-ray energy conversion efficiencies ranging from 1 to 10%. Figure S7 shows the results of four simulations with reduced peak density and larger foil separation compared to the simulations discussed in the manuscript. The initial transverse rms beam size is increased to $\sigma_\perp = 5\,\mu\mathrm{m}$, while the longitudinal size and charge are varied in each simulation from $1\,\mu\mathrm{m}$ to $5\,\mu\mathrm{m}$ and from 0.5 nC to 2 nC, respectively. The initial beam energy and normalized transverse emittance are 1 GeV and 3 mm mrad, respectively. This leads to a decrease in initial peak charge densities of approximately two orders of magnitude compared to those considered in the manuscript. Additionally, the interfoil separation is increased to $100\,\mu\mathrm{m}$ to account for the initially larger beam size. The simulations were performed using the particle-in-cell code CALDER-Circ [6] in a cylindrically symmetric geometry with a cell size of $\Delta_{r,z} = 0.5\,c/\omega_p$, where $\omega_p$ corresponds to that of $Al^{3+}$, and with four (eight) particles per cell for the electron beam ($Al^{3+}$ plasma).

In each case, the beam quickly focuses down to $\sim 1\,\mu\mathrm{m}$ rms beam size within $\sim 2\,\mathrm{mm}$ of propagation ($\sim 20$ foils) after which it starts radiating significant amounts of energy as gamma rays. After 16 mm of propagation the conversion efficiency exceeds the percent level in all four cases, and reach 10% for $\sigma_\perp = 5\,\mu\mathrm{m}$, $\sigma_\parallel = 1\,\mu\mathrm{m}$ and 2 nC [see Fig. S7(a)].

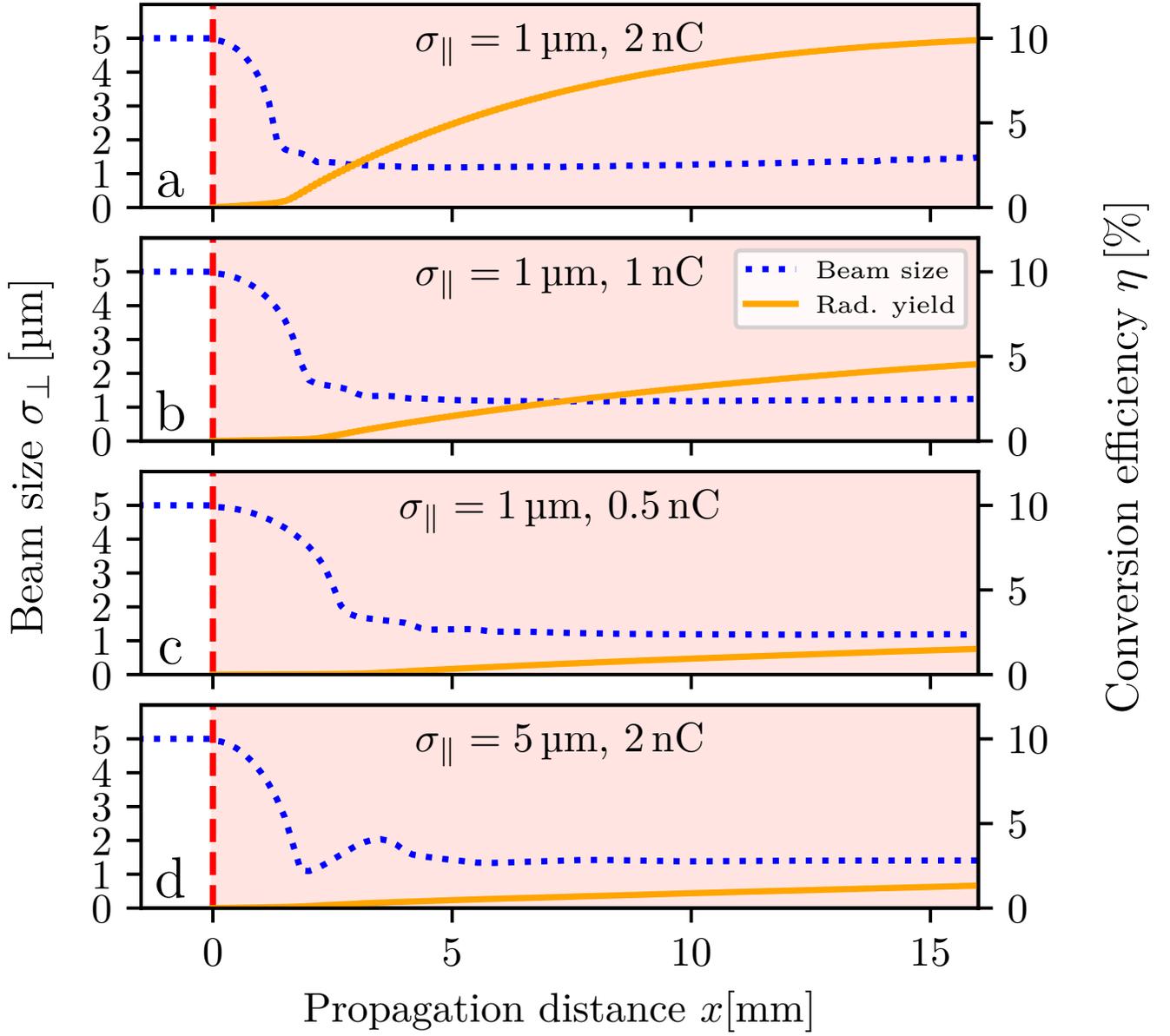

FIG. S7. Evolution of the beam transverse rms size (blue dotted line) and of the radiated energy (orange line) during propagation through a stack of $0.5\mu$m-thick Al foils with $100\mu$m interfoil spacing (red shaded area), starting at $x = 0$ mm (red dashed line). (a) The beam has $1(x) \times 5(y) \times 5(z)\,\mu$m rms size, 2 nC charge and 10 GeV energy. (b)-(c) Same as in (a) but 1 nC and 0.5 nC charge, respectively. (d) Same as in (a) but $5(x) \times 5(y) \times 5(z)\,\mu$m rms size.



\end{document}